\documentclass[12pt,a4paper]{article}
\usepackage{amsmath}
\usepackage{amsfonts}
\usepackage{amssymb}
\usepackage[active]{srcltx} 

\title{Quantum epistemology from subquantum ontology: quantum mechanics from theory of classical random fields}

\author{Andrei Khrennikov \\
  International Center for Mathematical Modeling \\in Physics, Engineering, Economics, and Cognitive Science\\
  Linnaeus University, V\"axj\"o, Sweden}

\begin{document}
\maketitle

\abstract{The scientific methodology based on two  descriptive levels, ontic (reality as it is ) and epistemic (observational), is briefly presented. Following Schr\"odinger,
we point to the possible gap between  these two descriptions. Our main aim is to show that, although ontic entities may be unaccessible for observations, they can be useful
for clarification of the physical nature of operational epistemic  entities.  We illustrate this thesis by the concrete example: starting with the concrete ontic model preceding quantum mechanics 
(the latter is treated as an epistemic model), namely, prequantum classical statistical field theory (PCSFT), we propose the natural physical interpretation for the basic quantum mechanical entity - 
the quantum state (``wave function'').   The correspondence PCSFT $\mapsto$ QM is not straightforward, it couples the covariance operators of classical (prequantum) random fields with 
the quantum density operators. We use this correspondence to clarify the physical meaning of the pure quantum state and the superposition principle - by using the formalism of classical 
field correlations. }

 \section{Introduction}
  
 Recently I presented \cite{AFB} analysis of the consequences of the final loophole free Bell's tests \cite{E1} - \cite{E3} based on the 
ontic-epistemic approach to physical theories, cf. Atmanspacher et al. \cite{H1}-\cite{H4}. This approach\footnote{ It has its roots 
in the old Bild conception elaborated by Hertz, Boltzmann, Einstein, Schr\"odinger, see, e.g.,   D'Agostino \cite{SCH_A1} for a good introduction.} prevents us from mixing two 
descriptive levels: ontic, ``reality  as it is'', and epistemic, representing results of  observations.\footnote{
In particular, in \cite{AFB} Bell's argument \cite{B1}, \cite{B2} was presented as the conjecture about {\it identification of the ontic 
states with the epistemic states.} From this ontic-epistemic viewpoint, the final loophole free test means that 
this conjecture about ontic-epistemic identification has to be rejected and the correspondence between the ontic and epistemic descriptions 
is not so straightforward as it was assumed in the ``Bell theorem''. }

Moreover, in \cite{AFB} I was very sympathetic to Schr\"odinger's viewpoint \cite{SCH_A2} that 
{\footnotesize\em theory and observations are not necessarily related in a term-to-term correspondence and a certain degree of independence 
exists between them.} In this paper I want to illustrate this Schr\"odinger's statement by the concrete example, the correspondence 
between the concrete ontic  model, {\it prequantum classical statistical field theory} (PCSFT), e.g.,  \cite{KH1}-\cite{Beyond} , and 
quantum mechanics (QM).

As respond to \cite{AFB}, I received a few messages stating that it is meaningless to consider ``fuzzy correspondence rules'' between subquantum and quantum models, 
since such considerations have no value for physics. I disagree with such claims. By the example of the PCSFT $\mapsto$ QM correspondence it will be shown that a  hidden ontic structure 
can clarify the real physical meaning of formal operational entities of the quantum formalism. Thus ontology can clarify the physical 
meaning of the basic epistemological  structures.

 In QM a quantum state $\psi$ is one of such main structures and its interpretation is characterized by huge diversity 
which is definitely a sign of theory's crises.  The main message of PCSFT  (the ontic model under consideration) to QM is that a quantum state is simply a 
normalized covariance operator of  a {\it ``prequantum'' random field,} a physical random field
propagating causally in space-time \cite{Beyond}.  Thus, in particular,  $\psi$ encodes not waves, neither physical
a la de Broglie \cite{DB}, \cite{Lochak}, Schr\"odinger \cite{SCH},  Einstein and Infeld \cite{EI}  nor probabilistic 
a la Born (see von Neumann \cite{VN} for detailed presentation), but correlations inside a random signal sent by a source of a prequantum random field (a source of quantum systems in the epistemic terminology).  

We stress that random fields (elements of the ontic model) represented by quantum pure states (elements of the epistemic model) are very special - they are concentrated in one 
dimensional subspaces of the $L_2$-space. (Quantum mixed states given by density operators represent random fields smashed over $L_2$-space.)

We also analyze the ontic structure behind the quantum notion of {\it superposition of pure states.} From the ontic viewpoint, creation of a superposition corresponds to fine tuning 
of signals represented by components of the superposition. Such turning has to generate  from one dimensional components a new one dimensional field.  We shall 
prove that in the probabilistic  terms this is equivalent to {\it  maximal correlation between components of the ``superposition random field''.}

The paper is competed by the appendix which has no direct relation to our main aim  (which is to illustrate that an  ontic (prequantum) model can clarify the (epistemic)   
quantum model). In the appendix we present another epistemic model generated by PCSFT describing observations of random signals with the aid of threshold detectors
and reproducing the probability distributions predicted by conventional QM \cite{KHD1}, \cite{KHD2} \cite{Beyond}. This is also an epistemic model (as the conventional quantum model). Thus the same ontological 
model can generated a variety of epistemic models for observed data. 

Section \ref{LA1} of the paper is devoted to the ontic-epistemic methodology of physical studies and its connection with the old Bild conception in 
modeling of physical phenomena; section \ref{LA2} is devoted to mathematical preliminaries about random physical fields and their 
correlations. Then in section \ref{LA3} we present PCSFT as an ontic model and QM as an epistemic model; correspondence between these models
(between the two descriptive levels) is established in section \ref{L}.  In section \ref{LAt} PCSFT is presented not in terms of random fields, but 
probability distributions on the $L_2$-space of physical fields (with the corresponding ontic model);  in section \ref{RS} we consider classical random field 
interpretation of pure quantum states and the principle of superposition as well as decoherence. This paper has an appendix which does not serve its main aim -
clarification of the physical content of quantum mechanical entities from PCSFT (the ontic subquantum model). In appendix we show that creation of 
an ontic model can lead to creation of new epistemic models; in our case different from QM.

\section{Ontic and epistemic descriptions of natural phenomena}
\label{LA1}

We discuss briefly   {\it the ontic-epistemic viewpoint} on natural phenomena,
established in quantum foundations by  Atmanspacher et al.  \cite{H1}-\cite{H4}.

There are ontic states, assigned to physical systems as ``they are'', and epistemic states representing knowledge that observers gain from measurements
on physical systems. {\it QM is about epistemic states.}\footnote{This viewpoint on quantum theory is strongly supported by information derivations of the 
quantum formalism, see, e.g., \cite{Z0}-\cite{Z1}, \cite{Ch1}-\cite{D2}, \cite{HD1}-\cite{HD3} as well as the subjective probability approach to QM known 
as Quantum Bayesianism (QBism) \cite{F1}-\cite{FX}.} This is in complete agreement with the Copenhagen interpretation of QM, especially Bohr's views.\footnote{The position 
of Bohr with respect to existence of ontic states is a more complicated issue. His view on this evolved. 
His final position was that ontic states of any kind are strictly forbidden in his
interpretation either before, during, or after measurements,  see Plotnitsky \cite{PL1}, \cite{PL3} for details.} 

We present the citation from the paper of Atmanspacher et al. \cite{H1}, p.  53 :

{\footnotesize\em ``Ontic states describe all properties of a physical system exhaustively.
(``Exhaustive'' in this context means that an ontic state
is ``precisely the way it is'', without any reference to epistemic
knowledge or ignorance.) Ontic states are the referents of individual
descriptions, the properties of the system are treated as
intrinsic properties. Their temporal evolution (dynamics) is reversible
and follows universal, deterministic laws. As a rule, ontic \index{ontic state} \index{epistemic state}
states in this sense are empirically inaccessible. Epistemic states
describe our (usually non-exhaustive) knowledge of the properties
of a physical system, i.e. based on a finite partition of the
relevant phase space. The referents of statistical descriptions are
epistemic states, the properties of the system are treated as contextual
properties. Their temporal evolution (dynamics) typically
follows phenomenological, irreversible laws. Epistemic states are,
at least in principle, empirically accessible.''} 

In classical mechanics the phase space description can be considered as the ontic description, 
here states are given by points $\lambda= (x,p)$ of phase space. The dynamics of the ontic state  is given by 
the system of Hamiltonian equations. 

We can also consider probability distributions on the phase space (or equivalently 
random variables valued in it). We call them {\it probabilistic ontic states.} Dynamics of 
probabilistic ontic states is given by the Louisville equation.

In classical physics we can (at least in principle) measure both the coordinate and momentum 
and hence ontic states can be treated as epistemic states as well (or it is better to say that here 
epistemic states can be treated as ontic states). Probabilistic ontic states represent probabilities for outcomes 
of joint measurement of position and momentum.

However, this was a very special, although very important, example of description of physical phenomena.
In general there are no reason to expect that properties of ontic states are approachable through our measurements.
There is a gap between ontic and epistemic descriptions, cf. also with `t Hooft \cite{TH1}, \cite{THX} and {G}  G. Groessing et al. \cite{G}.
In general the presence of such a gap also implies unapproachability  
of the probabilistic ontic states, i.e., probability distributions on the space of ontic states. De Broglie \cite{DB} called
such probability distributions {\it hidden probabilities} and distinguished them sharply from probability distributions
of  measurements outcomes, see also Lochak \cite{Lochak}.  (The latter distributions are described by the quantum formalism.)

This ontic-epistemic approach based on the combination of two descriptive levels for natural phenomena is closely related 
to the old {\it Bild conception} which was originated in the works of Hertz. Later it was heavily explored by Schr\"odinger  
in the quantum domain, see, e.g.,  \cite{SCH_A1},  \cite{SCH_A2} 
for detailed analysis. By Hertz one cannot expect to construct a complete theoretical model based explicitly on observable 
quantities. The complete theoretical model can contain quantities which are unapproachable for external measurement inspection.
For example, Hertz by trying to create a mechanical model for Maxwell's electromagnetism invented hidden masses.  The main 
distinguishing property of a theoretical model (in contrast to an observational model) is the {\it continuity of description}, i.e., 
the absence of gaps in description. From this viewpoint, the quantum mechanical description is not continuous: there is a gap 
between premeasurement dynamics and the measurement outcome. QM cannot say anything what happens in the process of measurement,
this is the well known measurement problem of QM \cite{VN}, cf. \cite{Opus}, \cite{Opusculo}. Continuity of description is closely related to causality. However, here 
we cannot go in more detail, see \cite{SCH_A1},  \cite{SCH_A2}. 

The important question is about interrelation between two levels of description, ontic-epistemic (or theoretical-observational).
In introduction we have already cited Schr\"odinger who emphasized possible complexity of this interrelation. In particular, 
in general there is no reason to expect straightforward coupling of the form, cf. \cite{B1}, \cite{B2}:
\begin{equation}
\label{OE}
\lambda \to a(\lambda),
\end{equation} 
where $\lambda$ is the ontic state (``hidden variable'') and $a$ is observable quantity. 

Since an epistemic description is typically probabilistic, it is natural to expect some form of correspondence between ontic probabilistic states 
and epistemic states. We shall explore this possibility in this note. 

Finally, we remark that the essence of the Bild conception is that the aim of physicists is creation of {\it mental images of natural 
phenomena} (the English translation of  German ``Bild'' is ``image''). Thus a complete and continuous (and typically causal) theoretical 
model is a mental image of reality as it is. (In the same way the ontic model is still a mental model of reality, i.e., even the ontic description 
is written by humans and not by nature.)  It is clear that our mind can generate 
a variety of mental images of the same natural phenomena, the same observational (epistemic) model can have a variety of theoretical (ontic) ``preimages''.
It is clear as well that the same observational data can be described by a variety of epistemic models. In particular, QM is just one of possible epistemic models
for observations over a special class of physical systems.  

Now we turn to the main objection to  creation of ontic models: {\it  Since ontic entities are unapproachable by our measurement devices, it is totally meaningless
to put efforts to theorizing which does not lead to explicit experimental consequences.}\footnote{I remember that one of the pioneers of development of 
quantum computer technology in Russia academician K.A. Valiev asked me in one intimate conversation: 
{\footnotesize\em ``Andrei, you are a smart man, why do you sacrifice your life by working on all these subquantum models? Now you can  make a real contribution to 
quantum information theory.''} (It was year 2000).} In short the answer is that by gaining the knowledge about nature we use not only outputs of observations 
(including our senses), but even the power of our mind, both logical and transcendental reasoning. This can lead us to laws of nature which are hidden in observational 
data. To connect observable quantities, we may need to invent ``hidden variables'', similar to Hertz hidden masses. The latter is may be not the best example, since 
the great project of the mechanical reformulation of Maxwell's electromagnetism   was not  completed - Hertz suddenly died. May be a better example is statistical thermodynamics
created, in particular, by another enthusiastic supporter of the Bild conception, by L. Boltzmann.\footnote{Bild conceptions of Hertz and Boltzmann were not identical, but here we do not 
have a possibility to discuss this point in more detail \cite{SCH_A1},  \cite{SCH_A2}. Of course, L. Boltzmann was strongly influenced by works of H. Hertz.}  At Boltzmann's time the position and momentum  of a molecular
or an atom were not approachable with the aid of existed measurement technology. (Moreover, even their existence was questionable, by, e.g., E. Mach.) And the great success
of statistical thermodynamics was a consequence of invention and use of these ``hidden variables''. (We remark that Boltzmann was agressivelly atacked not only by 
Mach, but by many prominent physicists and especially chemists, who claimed that his studies are not about physics, but metaphysics.) This example of the great success in physics
based on the creation of a theoretical (ontic) model also teaches us that it may (but need not) happen that development of technology can ``lift'' the ontic level 
to the  epistemic level. For a moment, we do not know where this may happen in quantum domain...  
 
\section{Classical random field and its covariance operator}
\label{LA2}

Let $H$ be a complex Hilbert space. A {\it random field} is a random variable valued in $H.$ 
Hence there is a probability space, i.e., a tripe 
${\cal P}=(\Omega, {\cal F}, P),$ where $\Omega$ is the space of elementary events, ${\cal F}$ is a family of its subsets representing events 
and $P$ is a probability measure on ${\cal F}.$ A random variable $\phi=\phi(\omega)$ is a map $\phi: \Omega \to H$ satisfying some restriction 
(measurability). 

The terminology ``random field'' is rooted to the case $H=L_2,$ where $L_2$ is the space of complex valued square integrable 
functions, e.g., on $\mathbf R^3.$ Here, for each chance parameter $\omega \in \Omega,$ the vector $\phi(\omega)$ is an element of $L_2.$ Hence,
it can be considered as a field, i.e., we have  the random field $\phi(\omega, x),$ where $x\in \mathbf R^3$ is the spatial variable.\footnote{Physicists do not like 
to use the chance variable $\omega$ in formulas. The presence of $\omega$ in coming considerations is the mathematical tradition.}      Sometimes, when the analogy 
with classical signal theory will be especially natural, we shall also use the terminology {\it ``random signal''}, see section \ref{RS}.

From the very beginning, we underline that operation in {\it complex} Hilbert space has no direct relation to QM; signal  theory (and, in particular, radio-physics)  
is often represented in the complex form; for example, we remind the Riemann- Silberstein representation of the classical electromagnetic field 
$\phi(x)= E(x) + i B(x).$ 

For a random field $\omega \to \phi(\omega),$ its average is defined as the vector $\bar{\phi} \in H$ such that
$$
\langle  \bar{\phi} \vert x \rangle= E  \langle \phi(\omega) \vert x \rangle= \int_\Omega \langle \phi(\omega) \vert x  \rangle d P(\omega),  \; x \in H.
$$
In this paper (and in the PCSFT-framework) we consider only random fields having zero average.

The {\it covariance operator} $B_\phi$ of  a random field $\phi= \phi(\omega)$ is defined by its bilinear form: 
$$
\langle  x_1\vert B_\phi\vert x_2 \rangle = E  \langle x_1 \vert \phi(\omega) \rangle  \langle \phi(\omega) \vert x_2 \rangle
$$
$$
= \int_\Omega \langle x_1 \vert \phi(\omega) \rangle \langle \phi(\omega) \vert x_2 \rangle d P(\omega),  \; x_1, x_2  \in H.
$$ 
(This is the complex covariance operator. We remark $H$ can be also considered as the real Hilbert space and the corresponding covariance operator is typically used in 
measure theory, see \cite{Beyond} 
for details.)

To make all previous operations justified, the dispersion of a random field has to be finite:
$$
\sigma^2_\phi= E \Vert \phi(\omega)\Vert^2= \int_\Omega  \Vert \phi(\omega)\Vert^2  d P(\omega) < \infty .
$$ 
Denote the class of such random fields, i.e., {\it  having zero average and finite dispersion,} by the symbol ${\cal R}.$ 

We point to the following formula playing the important  role in establishing ontic$\mapsto$epistemic correspondence. 
$$
\sigma^2_\phi= \rm{Tr} B_\phi.
$$ 

\medskip

For $\phi \in {\cal R},$ we list the basic properties of a covariance operator. It is 
\begin{enumerate}
\item Hermitian; 
\item Positive semi-definite;
\item Trace class.
\end{enumerate}
We also remark that any operator $B$ satisfying 1)-3)  can be considered as the covariance operator of a random field; for example, 
the Gaussian random field $\phi \sim N(0, B).$  Denote the class of operators satisfying 1)-3) by the symbol ${\cal C}.$ 

\medskip

We stress that a random field is not uniquely determined by its covariance operator (even if its average is fixed), thus the map 
$$ 
\mathfrak{c}:  {\cal R} \to {\cal C}, \phi \to B_\phi
$$
is not one-to-one. However, by restricting considerations to so called complex Gaussian random fields \cite{Beyond} we obtain the one-to-one correspondence. 
Denote the class of complex Gaussian random fields by the symbol  ${\cal RG}.$ Then the restriction of the map  $\mathfrak{c}$ on ${\cal RG}$
is one-to-one.

\section{States}
\label{LA3}

\subsection{Ontic states}

In PCSFT the ontic states of physical systems are ``physical fields''; mathematically they are represented by elements of 
$L_2=L_2(\mathbf R^3)$ space, $x \to \phi(x),$ and 
$$
\Vert \phi \Vert^2 = \int_{\mathbf R^3} \vert \phi(x) \vert^2 dx < \infty.
$$ 

One of the main characteristics of the ontic state is the possibility 
to assign to it the concrete energy. Here energy is treated totally classically, as in classical field theory.\footnote{For example, consider the classical electromagnetic 
field in Riemann-Silberstein representation, $\phi(x) = E(x) + i B(x).$ Then its energy density is defined as ${\cal E}_\phi (x)= E^2(x) + B^2(x)= \vert \phi(x) \vert^2.$}  
Each physical field, $x \to \phi (x),$  has the concrete 
energy  density
\begin{equation}
\label{L0}
{\cal E}_\phi (x)= \vert \phi(x) \vert^2, \; \; x \in \mathbf R^3.
\end{equation}
Its total energy is given by 
\begin{equation}
\label{L1}
{\cal E}_\phi= \int_{\mathbf R^3} {\cal E}_\phi(x) dx = \Vert \phi \Vert^2.
\end{equation}

Probabilistic ontic states (which are of the main interest for us) are given by probability distributions on $L_2$ or, as it is more convenient for our further
considerations, by random fields, i.e., random variables valued in $L_2.$  Such random variable can be expressed as a function of two variables, $\omega$ - the chance 
variable and $x$  - the space variable, $\phi= \phi(\omega, x).$ 

For a  random field $\phi(\omega, x),$ the quantities (\ref{L0}), (\ref{L1}) also depend on the chance parameter $\omega:  {\cal E}_\phi (x, \omega)$ and ${\cal E}_\phi(\omega).$    
Of course, these quantities fluctuate.  We can easily find the average of total energy which, in fact, coincides with the dispersion of the random field:
\begin{equation}
\label{L2}
\bar{\cal E}_\phi= E {\cal E}_\phi(\omega)= E \Vert \phi(\omega) \Vert^2.
\end{equation}
The problem of averaging the energy density is more complicated and we shall turn to it later.

This scheme is easily extended to the case of the abstract Hilbert space $H.$  
Here ontic states are vectors (in general non-normalized) belonging $H.$
Instead of the energy  density,  ``energy in the fixed point of space'', we can consider the ``energy along some direction'' $e \in H,$
\begin{equation}
\label{L0a}
{\cal E}_\phi(e)= \vert \langle \phi \vert e \rangle \vert^2. 
\end{equation}
Probabilistic ontic states are $H$-valued random variables, i.e., random $H$-fields. For such a state, we can find  the average of its energy  along the direction $e:$
\begin{equation}
\label{88}
E{\cal E}_\phi(e, \omega)= E \vert \langle \phi(\omega)\vert e \rangle \vert^2 =\langle e\vert B_\phi \vert  e \rangle.
\end{equation}
In the case of $H=L_2$ we may try to define the average of energy  density,  the ``average energy at the fixed point'' $x_0$  of physical space $\mathbf R^3.$ Formally, in $L_2$ 
points of space are represented by  the corresponding Dirac $\delta$-functions, i.e., $e_{x_0} (x)= \delta(x-x_0).$ The formula (\ref{88}) takes the 
form: 
\begin{equation}
\label{89}
E{\cal E}_\phi(x_0, \omega)= E \vert \langle \phi(\omega)\vert e_{x_0} \rangle \vert^2 =\langle e_{x_0}\vert B_\phi \vert  e_{x_0} \rangle.
\end{equation}
However, in general the latter quantity is not well defined, because the covariance operator $B_\phi$ of an arbitrary $L_2$-valued random field cannot be applied
to Dirac $\delta$-functions. Thus {\it the average of field's  energy density is not well defined.} (Of course, if field's realizations are smooth enough, e.g., they belong to the space of test functions, 
then  it can happen that the expression $\langle e_{x_0}\vert B_\phi \vert  e_{x_0} \rangle$ is well defined.) 

\medskip

We now turn the abstract framework. Here the total energy is also given by the squared norm. By selecting some orthonormal basis $(\vert e_k \rangle)$ in $H$ we can represent the total energy as the sum 
of energies emitted in the directions corresponding to the basis vectors:
\begin{equation}
\label{L1a}
{\cal E}_\phi(\omega)= \sum_k {\cal E}_\phi(\omega, e_k)=  \sum_k \langle e_k \vert B_\phi \vert  e_k \rangle=  {\rm Tr} B_\phi.
\end{equation}

Finally, we remark that for $H=L_2$ selection of some direction $e\in L_2$ can 
be engineeringly treated as selection of the direction of an antenna. 

\subsection{Epistemic states}

In QM -  an observational (epistemic) model - states are represented by density operators; denote the set  of such operators by the symbol ${\cal D}.$ 
Let $\rho \in {\cal D}.$ We list its  properties:
 \begin{enumerate}
\item Hermitian;   
\item Positive semi-definite;
\item Trace class; 
\item Trace equals to one.
\end{enumerate}
The reader can see that the lists of properties determining the classes of covariance operators of random fields  and density operators, ${\cal C}$ and ${\cal D},$ 
differ only by the additional fourth property - the unit trace.  (We remark that the difference due to 4) is not of a big value, because trace-one condition is simply 
the normalization constraint.) 
And a curious reader would definitely ask himself: What is the source of this coincidence? Is it just peculiar random coincidence? 
I think that this coincidence is not the result of the game of chance, but the consequence of the deep connection between theory of random fields and  QM.
In any event, this coincidence gives us the possibility to establish the natural correspondence between the ontic model, PCSFT, and the epistemic model, QM, see section \ref{L}.

\section{Ontic $\mapsto$ epistemic correspondence} 
\label{L}

The map from the space of probabilistic ontic states ${\cal R}$ to the space of epistemic states ${\cal D}$ is defined as
\begin{equation}
\label{L2pr}
J: {\cal R} \to {\cal D},  \rho_\phi= J(\phi)= \frac{1}{\sigma^2_ \phi} B_\phi.
\end{equation}

This map is not one-to-one, but  it maps ${\cal R}$ onto ${\cal D},$ i.e., each epistemic state is the image of some class of probabilistic
ontic states. This map induces the equivalence relation on the space of probabilistic ontic states, $\phi_1 \sim \phi_2,$ if $\rho_{\phi_1}= \rho_{\phi_2}.$
A factor-class is composed of random fields having the same covariance operator up to scalar factor - the average energy of the random 
field.  Such a factorization is the basic feature of ontic $\mapsto$ epistemic correspondence, cf. `t Hooft \cite{TH1}, \cite{THX} and author's works on so-called 
prespace \cite{KHR_PS}. Each factor-class ${\cal R}_\rho$ is huge; even by ignoring the energy-normalization we have a variety of random fields having the same 
covariance operator. However, by restricting probabilistic ontic states to (complex ) Gaussian random fields we can reduce degeneration to the energy-normalization. (And there 
can be presented some physical arguments that prequantum random fields are Gaussian.)  For $\phi_1, \phi_2 \in {\cal FG}, \phi_1 \sim \phi_2$ if and only if 
$\phi_1 =c \phi_2, c >0.$ Thus such random fields representing quantum systems in the same state $\rho$ differ only by the average energy.  
By selecting the random field $\phi_\rho$  with unit dispersion, i.e., unit average energy, we can represent other fields from this equivalent class 
as $u = \sigma_u \phi_\rho.$

One might say (and this is the typical comment on PCSFT):  {\footnotesize\em We are not able to detect prequantum random fields, for 
example, the components of electric and magnetic fields emitted by ``photon-sources'' (from the ontic viewpoint these are simply sources of 
classical weak electromagnetic-field).  Therefore the construction of the ontic PCSFT has no physical consequences.}

 On one hand, this is correct and the modern technology does not give us a possibility 
to perform measurements of PCSFT-quantities. On the other hand, technology develops quickly and we cannot exclude that soon we shall get such a possibility. 
In any event,  the correspondence (\ref{L2pr}) clarifies {\it the physical meaning of a quantum state as the normalized by the average energy covariance operator.} 
Thus the QM-formalism is about correlations. The correlation image of a quantum state is essentially more physical than the image of ``waves of probability''
used in the operational QM. Correlations of physical signals are well studied, e.g., in radio-physics; the ``waves of probability'' appear only in QM. Finally, we remark that 
the ``wave of probability''  is by itself a convenient notion, but one has to remember that this is the element of the epistemic description.

\section{Probability distribution of a random field and ontic $\mapsto$ epistemic correspondence}
\label{LAt}

In probability theory, for any random variable, we can consider its probability distribution. The same can be done 
for a random field. The reader should not be afraid that it is valued in (may be infinite dimensional) Hilbert space.
$$
p_\phi(O)= P(\omega \in \Omega: \phi (\omega) \in O),
$$
where $O$ is a ``sufficiently good subset'' of  $H$ (a Borel set).
Instead of random fields, one can operate with probabilities defined on $H.$  
In particular, as usual in probability theory, we can represent the average and covariance operator  in terms of the probability distribution 
of a random field:
$$
\langle \bar{\phi} \vert x \rangle= \int_H \langle u \vert x \rangle d p_\phi(u),  x \in H,
$$
and 
$$
\langle  x_1\vert B\vert x_2 \rangle = \int_H \langle x_1 \vert u \rangle \langle u \vert x_2 \rangle d p_\phi(u),  x_1, x_2 \in H,
$$
and, in particular, 
$$
\sigma^2_\phi= \int_H  \Vert u \Vert^2 d p_\phi(u) .
$$ 
Denote the space of probability measures on $H$ with zero average and finite dispersion by the symbol ${\cal M}.$ 

\medskip

The space ${\cal M}$ can be used as well as the space of probabilistic ontic states.  The ontic $\mapsto$ epistemic correspondence 
is established by the following mapping:  
\begin{equation}
\label{L2pr}
J: {\cal M} \to {\cal D},  \rho_p= J(p)= \frac{1}{\sigma^2_ p} B_p,
\end{equation}
where $B_p$ denotes the covariance operator of the probability measure $p \in {\cal M}(H).$ 

\section{Pure states, the principle of superposition} 
\label{RS}

The space of quantum state ${\cal D}$ contains the very important subspace ${\cal D}_{\rm{pure}}$ consisting of pure states. They are represented by normalized vectors of $H.$

Now let $\psi$ be an arbitrary vector. Consider the one-dimensional linear space of vectors which are collinear to $\psi: L_\psi=\{\phi= c \psi: c \in \mathbf C\}$
and the operator  
\begin{equation}
\label{PROJ}
\pi_\psi = \vert \psi \rangle \langle \psi \vert,
\end{equation}
i.e., $\pi_\psi \phi = \langle \psi \vert \phi\rangle \vert \psi \rangle, \phi \in H,$ 
mapping $H$ onto $L_\psi.$ If $\psi$ is normalized, then $\pi_\psi$ is the  
orthogonal projector onto this subspace.  Any pure state $\psi \in H$ can be  treated as the density operator $\rho_\psi=
 \pi_\psi \in {\cal D}_{\rm{pure}}.$ Normalization of a pure state by the norm-one corresponds to the trace-one normalization 
of the density operator.

In the light of previously introduced ontic $\mapsto$ epistemic correspondence it is interesting to find ontic preimages of pure quantum states. 
In this way, i.e., by appealing to the ontic model behind QM, we can clarify the notion of a pure state and difference between pure and
mixed states. The modern state of art is characterized by diversity of viewpoints, up to statements that there are no pure states, see, e.g., 
\cite{Opus}, \cite{Opusculo}. Our final aim is to clarify the meaning of one of the basic principle of QM, {\it the principle of superposition:}   

{\it Let $\psi_k, k=1,...,n$ be pure quantum state. Then each their linear combinations (normalized by 1) is again a pure state.}

\medskip

First we present a few facts about quantum pure states which will be useful in further considerations. Let $H$ be two dimensional with the orthonormal basis
$(\vert e_1\rangle, \vert e_2\rangle).$ Here any pure state $\psi$ can be represented as the linear combination of the basis vectors, 
$\psi =c_1 \vert e_1\rangle + c_2 \vert e_2\rangle,$ where $\vert c_1\vert^2 + \vert c_2\vert^2=1.$ The corresponding density operator $\rho_\psi= \pi_\psi$ has the matrix 
\begin{equation}
\label{PROJ1}
 \left( \begin{array}{ll}
\vert c_1\vert^2  & c_1 \bar{c}_2\\
\bar{c}_1 c_2  & \vert c_2\vert^2
\end{array}
 \right ).
 \end{equation}
Now we present a simple mathematical fact about measures on Hilbert spaces \cite{Beyond}.

\medskip

{\bf Proposition 1.} {\it The covariance operator $B_p$ of the measure $p \in {\cal M}$ has the form  (\ref{PROJ}),
where $\psi$ is an arbitrary (nonzero) vector of $H,$ if and only if the measure $p$ is concentrated on the one-dimensional subspace $L_\psi.$  } 

\medskip

Thus any random field with the covariance operator of the form (\ref{PROJ}) is concentrated in the subspace $L_\psi$ of $H: \phi(\omega) \in L_\psi$ 
(for almost all $\omega$)  and vice versa. Thus {\it pure states correspond to random fields taking values in one dimensional subspaces.} 

From the ontic viewpoint, preparation of physical systems in pure states is fine turning of prequantum random fields to approach (with probability one) their
concentration in corresponding one dimensional subspaces. Here it is important that one dimensionality can be guarantied only with probability one; so for some 
$\omega,$ such a ``pure state field'' can leave $L_\psi,$ but the probability of such an event is negligibly small.     

\subsection{Maximally correlated signals}
 
Before to proceed to the ontic interpretation of the principle of superposition, we consider superposition (linear combination) of two classical signals. This illustrative 
example will be useful for clarification of the correlation-interpretation of this quantum principle. 

Consider two dimensional Hilbert space with the basis $(\vert e_1\rangle, \vert e_2\rangle)$ and two signals $\phi_1(\omega), \phi_2(\omega)$ valued in one dimensional 
subspaces corresponding to the basis vectors. As always, it is assumed that they have zero averages and finite dispersions. 
 We are interested in the covariance matrix of  the  sum of these signals, $\phi(\omega) = \phi_1(\omega) + \phi_2(\omega).$ Since signals $\phi_k(\omega), k=1,2,$
are collinear     to the basis vectors, we can represent them as $\phi_k(\omega)= \xi_k(\omega) \vert e_k\rangle,$ where $\xi_k(\omega)= \langle e_k \vert \phi_k(\omega) \rangle.$ 
We proceed with these scalar random variables.  First we study the case of real-valued variables: 
$\sigma_k^2= E \xi_k^2(\omega), \sigma_{12} = E \xi_1(\omega) \xi_2(\omega).$ The covariance matrix has the form:
\begin{equation}
\label{PROJ2}
B_\phi= \left( \begin{array}{ll}
 \sigma_1^2 &  \sigma_{12} \\
\sigma_{12}  & \sigma_2^2
\end{array}
 \right).
 \end{equation}
Now we ask the question: What are constraints on the covariance matrix guaranteeing concentration of the random variable $\phi$ in one dimensional 
subspace?  

By Proposition 1 it happens if and only if $B_\phi$ has the  form (\ref{PROJ}). By taking into account (\ref{PROJ1}) and remembering that we study the real case, we obtain that 
this happens if and only if the covariance-coefficient $\sigma_{12}$ is factorized as 
$\sigma_{12}= c_1 c_2,$ where $c_k= \pm \sigma_k.$  Thus we obtain that $\sigma_{12}= \pm \sigma_1 \sigma_2.$ Consider now the {\it correlation-coefficient}
$\rm{cor} \; (\phi_1, \phi_2) = \rm{cor} \; (\xi_1, \xi_2)= \frac{\sigma_{12}}{\sigma_1 \sigma_2}.$ We derived the following mini-proposition: 

\medskip   

{\bf Proposition 2R.} {\it The sum  of two one-dimensional orthogonal signals
is also one-dimensional if and only if  they are maximally (anti-)correlated: 
\begin{equation}
\label{PROJ3}
\rm{cor} \; (\phi_1, \phi_2) = \pm 1.
\end{equation}
If this case the sum is valued in $L_\psi,$ where 
\begin{equation}
\label{7p}
\psi=c_1 \vert e_1\rangle + c_2 \vert e_2\rangle
\end{equation}
 and 
$c_k= \pm \sigma_k.$}

\medskip
 
In the complex case $ \sigma_{12} = E \xi_1(\omega) \bar{\xi}_2(\omega).$ 
Consider the circles $S_k=\{z \in \mathbf C: \vert z\vert = \sigma_k\}.$ The covariance matrix (\ref{PROJ2}) of the signal $\phi(\omega)$ has the form (\ref{PROJ1}) if and only if 
there exist coefficients $c_k \in S_k$ such that $\sigma_{12}= c_1 \bar{c}_2.$ Thus we obtained the complex version of Proposition 2R:

\medskip

{\bf Proposition 2C.} {\it The sum  of two one-dimensional orthogonal signals
is also one-dimensional if and only if  they are maximally correlated:
\begin{equation}
\label{PROJ4}
\vert \rm{cor} \; (\phi_1, \phi_2) \vert = \pm 1.
\end{equation}
}

\medskip   
If (\ref {PROJ4}) holds, then the sum is valued in $L_\psi,$ see (\ref{7p})  and  $c_k\in S_k.$ We remark that the subspace $L_\psi$ does not depend on the choice 
of the ``factorization-coefficients''   $c_k\in S_k.$ 

In particular, if the sum of dispersions of signals equals to 1, i.e., $\sigma_1^2+ \sigma_2^2=1,$
 then $\Vert \psi \Vert^2 =\rm{Tr} \pi_\psi=1;$ hence $\psi$ is a pure state.  Thus {\it the quantum superposition  (\ref{7p}) of two (orthogonal) pure states 
$\psi_k= \vert e_k\rangle$ represents the sum of two maximally correlated signals which are colinear to these pure states.}

The previous study can be easily generalized to the case of $n$ random signals concentrated in one-dimensional orthogonal subspaces of $H.$ 
Let $(\vert e_j\rangle)$ be an arbitrary orthonormal basis in $H$ and let $\phi_j(\omega)$  be signals which are collinear to the basis vectors.   
Consider the signal 
\begin{equation}
\label{PROJ5Z}
\phi(\omega)= \sum_j      \phi_j(\omega) =  \sum_j     \xi_j(\omega)  \vert e_j\rangle.
\end{equation}
 In this basis its covariance matrix 
 $B_\phi=(b_{km}),$ where $b_{km}= E  \xi_k(\omega) \bar{\xi}_m(\omega);$ in particular, its diagonal elements coincide with dispersions of summands
$\sigma_k^2=  b_{kk}$ and its trace with the total dispersion of the signal $\phi(\omega).$  

Now we remind that, for a vector $\psi \in H,$ the matrix of the operator $\pi_\psi$ has the form $(c_k \bar{c}_m),$ where $\psi = \sum_k c_k \vert e_k\rangle.$ 
To be concentrated in one-dimensional subspace of $H,$ the signal $\phi$ has to have the covariance matrix $B_\phi$ of such a type, i.e., its elements has to satisfy 
the constraints: $b_{km}= c_k \bar{c}_m,$ where $c_k \in S_k.$ This means that the correlations between the component-signals have to be maximal, i.e.,  for all 
pairs  $(k,m),$ 
\begin{equation}
\label{PROJ5}
\vert \rm{cor} (\phi_k, \phi_m) \vert = 1. 
\end{equation}
In such a case the random field given by (\ref{PROJ5Z}) is valued in $L_\psi,$ where $\psi = \sum_k c_k \vert e_k\rangle,$ where $c_k \in S_k.$

Now we consider the ontic preimage of superposition of non-orthogonal states. Let $\phi(\omega) = \phi_1(\omega) + \phi_2(\omega),$
where $\phi_k(\omega)= \sum_j   \xi_{kj}(\omega)  \vert e_j\rangle,$ i.e., in general these signals are not orthogonal.  Then (by linear algebra) $\phi(\omega)=    \sum_j   (\xi_{1j}(\omega) + \xi_{2j}(\omega))  \vert e_j\rangle.$
Now the previous considerations can be used for the  random variables $\xi_{j}(\omega)= \xi_{1j}(\omega) + \xi_{2j}(\omega).$ And the random field $\phi(\omega)$ superposed
of random fields $\phi_k(\omega)$ representing (in the ontic model) some pure states $\psi_k$ (of the epistemic model) also represents a pure state if the correlations between 
coordinate-signals $\xi_{j}(\omega)$ are maximal, see   (\ref{PROJ5}).

\subsection{Ontic picture of superposition of quantum states; decoherence}

PCSFT provides the new  viewpoint on superposition of quantum states - by treating them as images of prequantum random fields.
Superposition of the form (\ref{PROJ5Z}) does not represent simply superposition of waves, neither the ``waves of probability'' (in the spirit of Born) 
nor the ``physical waves'' (in the spirit of Schr\"odinger and De Broglie).   It represents very fine mutual turning of random signals $\phi_j(\omega)$ 
represented by the quantum states $\psi_k= \vert e_k\rangle$ such that these signals are maximally correlated. The latter explains the  peculiarity  of creation and 
preservation of superpositions. These are processes of perfect turning of mutual correlations in random signals. By loosing correlations between some 
components of random signals (in the ontic model -PCSFT) we destroy  the state of superposition (in the epistemic model - QM). It is clear that it is more difficult to create and preserve 
maximal correlations between the components  of multi-component signal or (by using the epistemic terminology) to create and preserve multi-dimensional 
superpositions.   

Decoherence is one of the most complex notions of the epistemic model (QM) and it plays the fundamental role in variety important problems of quantum technology, e.g., 
quantum computing. From the ontic viewpoint decoherence is the process of destruction of the maximal correlations between components of a signal. From the classical 
signal viewpoint, it is clear that it is difficult to preserve maximal correlations for sufficiently long time. 
 
 \section{Concluding notes}
 
 We hope that in this paper we demonstrated that creation of ontic causal models for QM which is treated as the epistemic model describing measurements is not totally meaningless.
 Even if coupling between two descriptive levels, ontic and epistemic, is not straightforward we can extract some information from the ontic model which can be useful to establish 
 the proper interpretation of the basic quantities of the epistemic model. (We stress that precisely the absence of the ontic background for QM led to elaboration of numerous mutually 
contradictory interpretations of QM which can be considered as a sign of a crisis in quantum foundations). 

By appealing to the special ontic model of the classical random field type (PCSFT) we clarified the meaning of a quantum state (in particular, a pure quantum state - ``wave function'').
This appealing to PCSFT resolved the contradiction between interpretations based on ``physical waves" and ``waves of probability''. Yes, there are ``physical waves'', they are mathematically 
represented as classical random fields, e.g., photons are just classical electromagnetic fields. Such subquantum fields propagate causally in physical space. However, the wave function 
is not the ontic wave. The wave function and more generally the quantum state are covariance operators of subquantum random fields. Thus the wave function can be interpreted probabilistically 
as the ``wave of correlations.''

Another lesson is that {\it pure states really exist}, cf. \cite{Opus}, cite{Opusculo}. They are epistemic images of very special signals concentrated on one dimensional subspace of $L_2.$ 
By operating with superposition of pure quantum states we, in fact, operate with signals having {\it maximally correlated components.} This explains the difficulty of 
creation multi-dimensional superpositions of quantum states and clarify the process of decoherence. 

\section*{acknowledgments} This research is supported by the grant ``Mathematical Modeling of Complex Hierarchic Systems'', Faculty of Technology, Linnaeus University, 
and ``Quantum Bio-Informatics'', Tokyo University of Science. My interest to the two levels description approach to quantum foundations was ignited by H. Atmanspacher, but it took
more than 20 years to start to understand this scientific methodology.

\section*{Appendix: Detection model based on PCSFT}

By writing this paper I was in doubts whether to add this appendix to it or not... The aim of the paper was to show that even an ontic model 
having no straightforward relation to observations, in our case PCSFT,  can generate a nontrivial interpretational impact for the corresponding epistemic model, in our case QM. 
And I hope that in the main body of this paper this aim was approached. In this appendix I want to present another epistemic model corresponding to PCSFT.
(Yes, the same ontic model can generate a variety of epistemic models.)   
In this appendix we complete PCSFT, our ontic model, with a detection model exploring directly the random field structure of PCSFT (which is absent in the quantum epistemic model)
\cite{KHD1}, \cite{KHD2}.  
 
Since physical systems are represented by random fields,  we can proceed as in the usual signal theory. Observations over such fields are mathematically described 
through selections of orthonormal bases in $H$ (for physics, the case $H=L_2$ is of the main interest). Thus to fix an observable $A,$ we fix some concrete orthonormal 
basis $(\vert e_k\rangle).$ Then, again as in classical signal theory, we split a random signal $\phi= \phi(\omega)$ into the 
components $\xi_k(\omega)=\langle  e_k \vert  \phi(\omega)\rangle$ detected at channels $k=1,2,....$ 
(This spitting has nothing to do with the quantum theory; classical fields can be expended with respect to bases in $L_2.)$

 The probability of detection in the $k$th channel is proportional to the 
average energy of  the scalar signal $\xi_k(\omega):$ 
\begin{equation}
\label{PROJ6}
p_\phi (k)=\frac{E  \vert \xi_k(\omega)\vert^2}{\sum_j E  \vert \xi_(\omega)\vert^2} 
= \frac{\langle e_k\vert B_\phi\vert e_k \rangle}{{\rm Tr} B_\phi}={\rm Tr} \rho \vert e_k\rangle \langle e_k \vert,
 \end{equation}
 where $\rho=J(B_\phi)=\frac{1}{{\rm Tr} B_\phi} B_\phi.$ 
 This is nothing else than the Born rule which is postulated in the epistemic quantum model. We derived it from very natural detection rule: probability is proportional to 
 signal's energy.
 
 The main objection to this purely field approach to detection probabilities is that  a physical field by split  into a few channels should present in all of them 
and hence induce simultaneous detections. We start the discussion on this objection with the following provocative question: Does QM prevent simultaneous detections?
The answer is `yes' and `no' at the same time. If one consider QM as endowed with the  von Neumann-L\"uders projection postulate  and moreover treat ``collapse of the 
wave function'' as a physical event, then the answer to the above question is `yes'. The act of detection implies collapse and hence all components of the wave function, besides
one corresponding to the output of measurement, disappear. However,   the  von Neumann-L\"uders projection postulate is a very special postulate of QM and not all authors 
agree to consider it as belonging to the body of QM, cf.  \cite{BL}.\footnote{Even those who are fine with this postulate need not share the ``physical collapse'' viewpoint. Moreover, nowadays 
it is practically forgotten that personally von Neumann declared the possibility to use the projection postulate only for observables with non-degenerate spectrum.} 

It seems that in QM the only possibility to motivate impossibility of simultaneous detections is to consider the coefficient of second order coherence $g^{2}$ and show 
that 
\begin{equation}
\label{PROJ7}
g^{2}(0)<< 1.
\end{equation}
However, this is the statistical impossibility; simultaneous detections are possible, but the probability of such events is very small. 

We also point out that proceeding without physical  state collapse we cannot derive discreteness of detection events from the quantum formalism. This conclusion  
may be surprising, since it is generally assumed that quantum measurement theory is about discrete events of detection. But it seems that the relict of the collapse interpretation 
of the projection postulate.    
  
 Now we turn to our ontic model and detection theory (which is epistemic derivative of PCSFT - different from QM).  In PCSFT we proved 
that (\ref{PROJ7}) holds true. Moreover, by assuming that all detectors are of the threshold type we obtain discrete detection events: a detector 
``eats'' 
the prequantum field until it collects energy overcoming the detection threshold, then it clicks.   Events of simultaneous detections in a few channels are rare, see
((\ref{PROJ7})), see \cite{KHR_L1}, \cite{KHR_L2}.

\end{document}